\documentclass[11pt,a4paper]{article}
\usepackage[left=2.5cm,top=2.5cm,right=2.5cm,bottom=2.5cm]{geometry}
\usepackage{graphicx}
\usepackage{bm}
\usepackage{doi}
\usepackage{color}

\usepackage{comment}

\begin{document}

\clearpage
\noindent \textbf{\LARGE Nanoscale magnetic compasses}\\
\vspace{0cm}
\noindent \textbf{Hidetsugu Shiozawa,$^{1\ast}$ Desai Zhang,$^{2}$ Michael Eisterer,$^{3}$ Paola Ayala,$^{1}$ Thomas Pichler$^{1}$ Martha R. McCartney,$^{2}$ David J. Smith.$^{2}$} \\
\noindent $^{1}$Faculty of Physics, University of Vienna, Boltzmanngasse 5, 1090 Vienna, Austria\\
\noindent $^{2}$Department of Physics, Arizona State University, Tempe, AZ 85287-1504, USA\\
\noindent $^{3}$Atominstitut, TU Wien, Stadionallee 2, 1020 Vienna, Austria\\
\noindent $^\ast$To whom correspondence should be addressed; E-mail:  hidetsugu.shiozawa@univie.ac.at.\\
\vspace{5mm}

\begin{abstract}
We have synthesized nanoscale magnetic compasses with high yield. These ferromagnetic iron carbide nano-particles, which are encapsulated in a pair of parallel carbon needles, change their direction in response to an external magnetic field. Electron holography reveals magnetic fields confined to the vicinity of the bicone-shaped particles, which are composed of few ferromagnetic domains. Aligned magnetically and encapsulated in an acrylate polymer matrix, these nano-compasses exhibit anisotropic bulk magnetic permeability with an easy axis normal to the needle direction, that can be understood as a result of the anisotropic demagnetizing field of a non-spherical single-domain particle.
This novel material with orthogonal magnetic and structural axes could be highly useful as magnetic components in electromagnetic wave absorbent materials and magnetorheological fluids.
\end{abstract}

\section*{Introduction}
The magnetic compass could be considered as one of the more important historical discoveries that have changed the world. Dating back to ancient times, the first compasses were composed of lodestone, a naturally-magnetized iron ore, floating on water. Mechanical compasses have since become fundamental tools for marine navigation. Nowadays, a compass 'app' comes with every smartphone with an integrated electronic magnetometer.
While cutting-edge fabrication techniques allow mechanical compasses to be miniaturized as much as possible, their traditional role has been replaced in many cases by more compact and sophisticated technologies such as magnetic field sensors, gyroscopes and satellite navigation systems.

Here, we demonstrate a controlled and scalable synthesis of nanoscale compasses that could be implemented in modern technologies. Due to their unique properties \cite{Faraji10JotICS,Kodama99JoMaMM,Lu07ACE,Goya03JoAP,Gambardella03S,Rong06AM}, these nano-sized magnets are promising as magnetic components in many applications such as electromagnetic wave absorbers, magnetic recording, catalysts and biomedicine. Many scientific reports on magnetic nanoparticles published recently illustrate the growing trend towards biomedical applications \cite{Berry03JoPDP,Laurent08CR,Lewin00NB,Pankhurst03JoPDP,Sun08ADDR,Tartaj03JoPDP,Arruebo07NT,Gao09AoCR,Ito05JoBaB,Jordan99JoMaMM,Kim08ACE} including diagnostics, therapeutics, gene and drug delivery \cite{McBain08IJoN,Oh10B,Mendes13JoMCB}.
Core-shell heterogeneous structuring is the key to functionalizing magnetic nanoparticles. Interactions at the core-shell interface can lead to stable and excellent nanomagnets. A hydrophobic shell makes the particle bio-compatible while chemically inert shell materials such as graphite \cite{Bokhonov02JoAaC,Si03C,Ding06JoPCB,Cao09C,Bystrzejewski07BE,RUOFF93S,Lu05C,SAITO95C} and carbon nanotubes \cite{Setlur98JoMR,Liang00JoCG,Hampel06C,Briones-Leon13PRB} make them sustainable against environmental factors.

It was shown recently that high-pressure pyrolysis of ferrocene leads to the formation of iron carbide nanoparticles naturally encapsulated in moustache-shaped graphitic spiral shells \cite{Shiozawa11NL,Shiozawa13SR}. In the present work, we have optimized the original synthesis technique for needle-like shells and studied their magnetic properties on both microscopic and macroscopic scales. Using electron holography in a transmission electron microscope (TEM), we demonstrate that the cores are ferromagnetic while the shells are not. Bulk magnetization measurements reveal magnetic anisotropy of the core with an easy axis normal to the needle axis, that can be understood as a result of the anisotropic demagnetizing field of a bicone-shaped single-domain ferromagnet. This geometry explains the response of the nano-objects to a magnetic field, acting as compasses that allow them to be aligned magnetically within a polymer matrix. The uniquely coupled shape and magnetic anisotropy of the nano-compasses should be highly advantageous as magnetic components in radio wave/microwave absorbent materials and magnetorheological fluids.

\section*{Results and Discussion}
\begin{figure*}
    \centering
    \includegraphics[width=\textwidth]{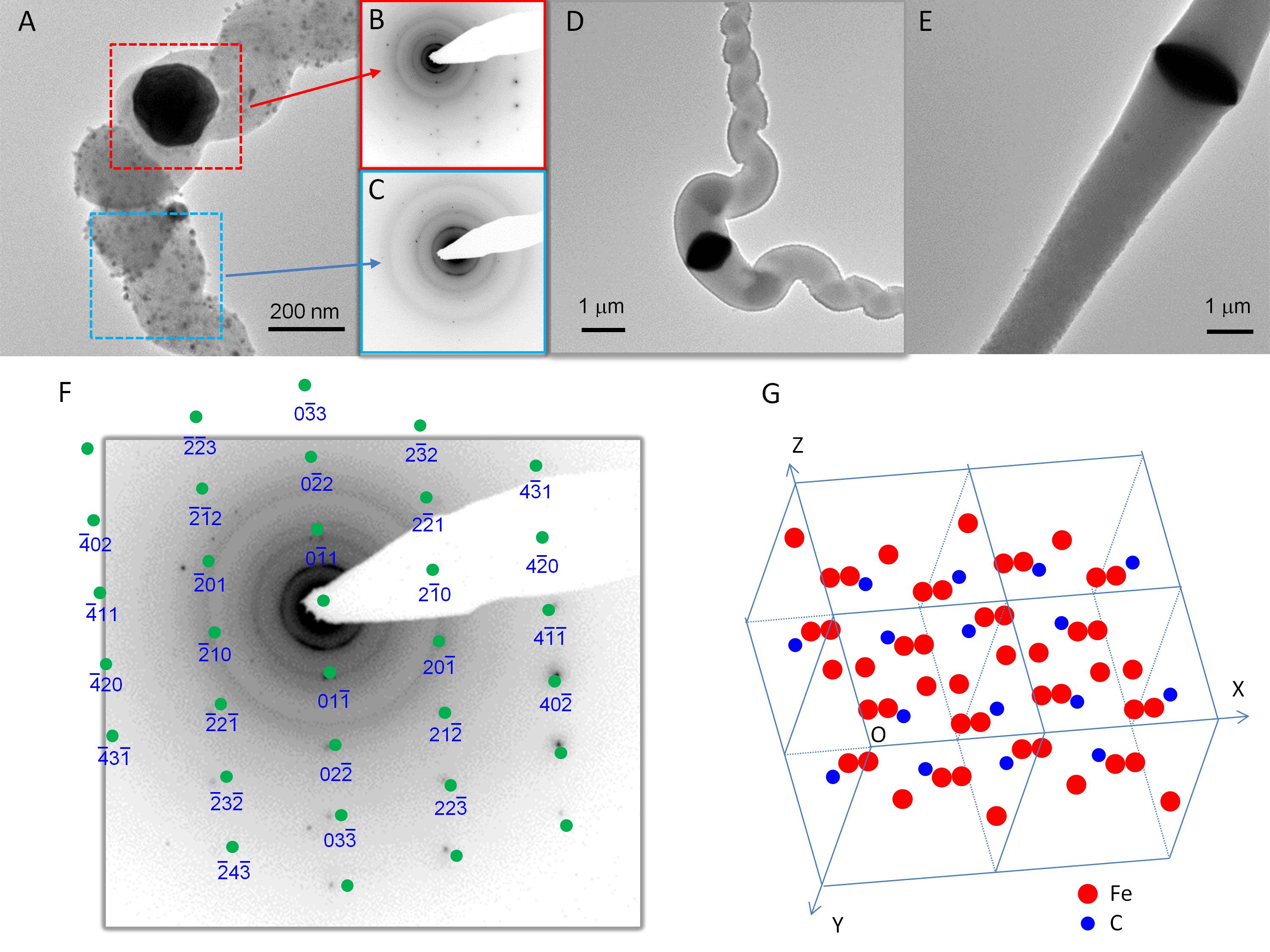}
    \caption{
    Transmission electron micrographs showing nano objects with three different sizes (\textbf{A}, \textbf{D} and \textbf{E}), with bright graphite arms and dark iron-carbide core. Selected-area electron diffraction (SAED) patterns collected within the dashed square shown in red (\textbf{B}) and in blue (\textbf{C}) in panel \textbf{A}.
    (\textbf{F}) Reciprocal lattice points for cementite oriented in the direction shown in panel \textbf{G}, superimposed onto the SAED pattern.
    \label{tem}}
\end{figure*}

Panels A, D and E in Figure~\ref{tem} are TEM micrographs of nano-objects with difference sizes, showing graphite arms with bright contrast and iron-carbide cores with dark contrast.
As previously reported, larger objects have straighter arms and flatter cores \cite{Shiozawa11NL,Shiozawa13SR}. The smallest object in panel A is sufficiently thin that a selected area electron diffraction (SAED) can be obtained through its core. The SAED pattern shown in B exhibits a geometry that matches a simulated pattern of cementite oriented in the direction, demonstrated in panels F and G. It is noteworthy that the crystal orientation with regard to the arms' axis is roughly the same as observed in previous work \cite{Shiozawa13SR}, indicating that the facetted crystal core extrudes the carbon arms into a low-index lattice direction.

\begin{figure*}
    \centering
    \includegraphics[width=\textwidth]{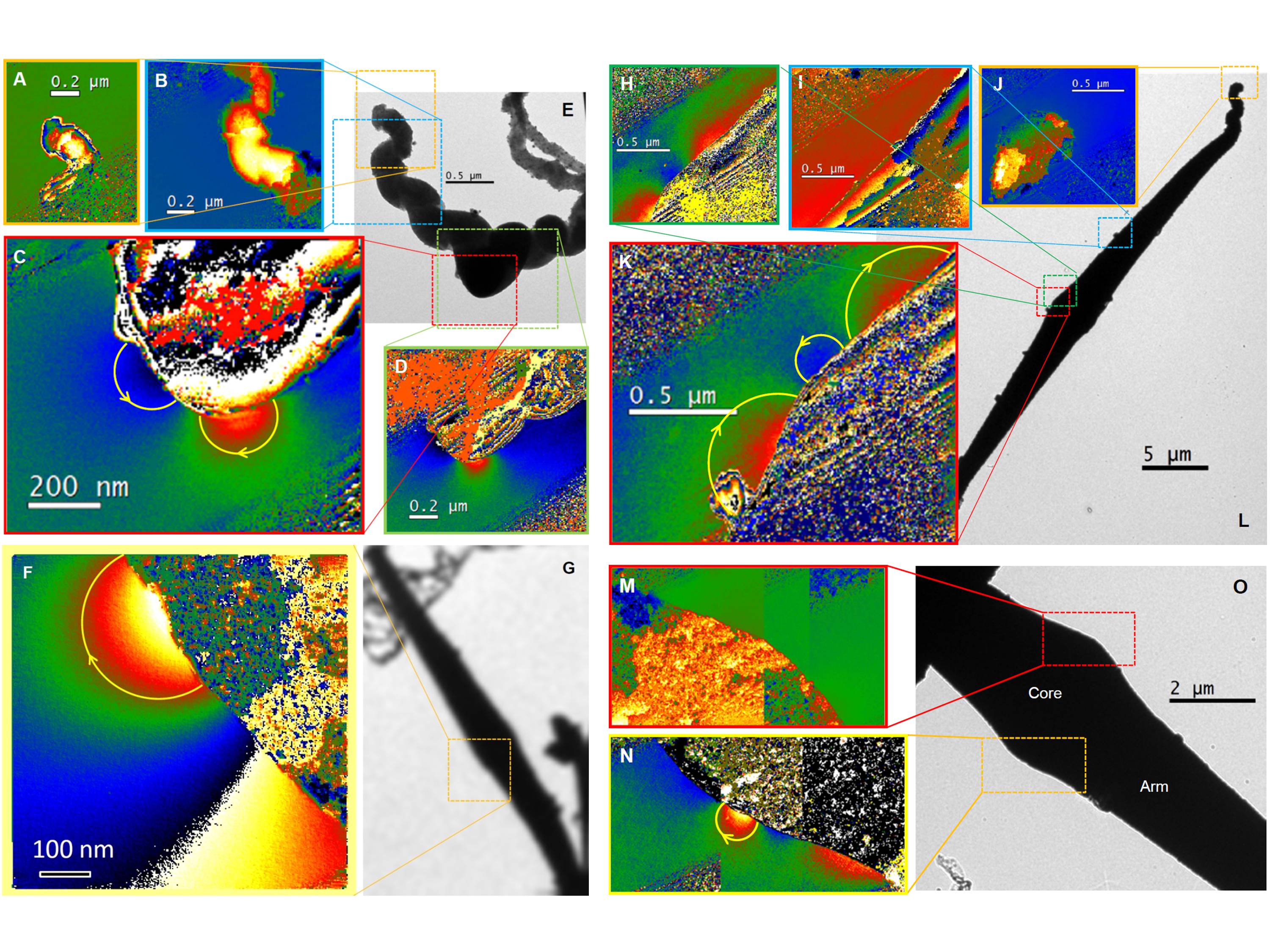}
    \caption{
    (\textbf{A}--\textbf{D}) Phase maps from reconstructed holograms across the small spiralling object shown in panel \textbf{E}.
    (\textbf{F}) Phase map collected in the vicinity of the core of the straight object shown in panel \textbf{G}.
    (\textbf{H}--\textbf{K}) Phase maps across the straight object collected within the dashed squares shown in panel \textbf{L}.
    (\textbf{M} and \textbf{N}) Phase maps in the vicinity of the core of the straight object in panel \textbf{O}, showing magnetic fields only on one side.
    The phase maps shown in panels \textbf{M} and \textbf{N} were assembled using three phase maps in order to reveal larger area.
    \label{phase}}
\end{figure*}

Electron holography is a TEM technique that can reveal phase shifts in electron waves caused by magnetic fields \cite{LICHTE1991,Tonomur92AiP}. Figure \ref{phase} shows reconstructed phase maps associated with a small spiralling object (A--D) and large straight nano-objects (F--O). The colour changes correspond to electron-wave phase shifts due to the magnetic fields. Here, the magnetic flux lines are observed in vacuum only in the vicinity of the iron-carbide core. Possible magnetic flux lines that can cause the phase distributions are illustrated by arrows in panel C, F, K and N. Apparently, the iron-carbide core is a ferromagnet consisting of a few magnetic domains. The uniform phase distribution around the tip of the carbon spiral (see panel A) as well as the equal phase on both sides of the carbon arm (panel B), indicate no magnetic ordering across the graphitic arms.

\begin{figure*}
    \centering
    \includegraphics[width=\textwidth]{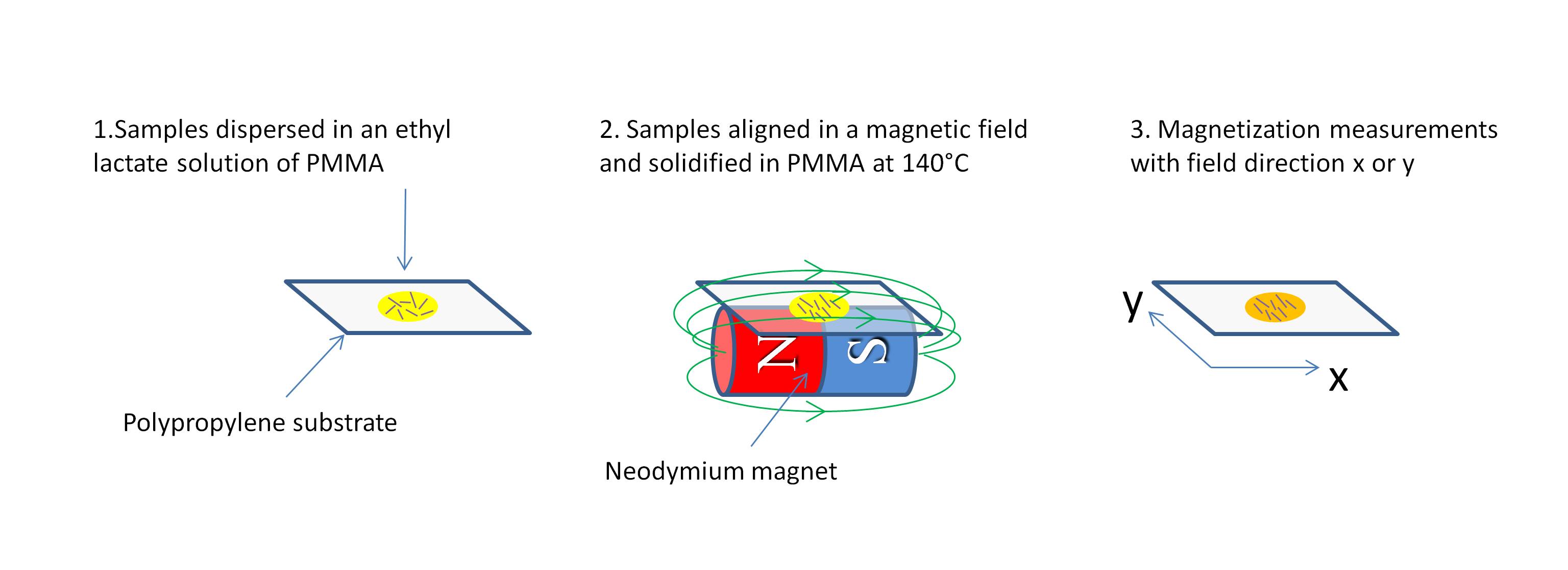}
    \caption{Encapsulation and alignment of nano-objects in PMMA.
    \label{pmma}}
\end{figure*}

A key question is how the magnetically heterogeneous nano-objects respond to applied magnetic fields individually and collectively, which is essential as far as potential applications are concerned. In order to answer this question, the synthesis was optimized for straight needle-like objects that were then dispersed in an ethyl lactate solution of polymethyl methacrylate (PMMA). Observations under an optical microscope reveal that needle objects trace a magnetic field. The Supplementary Video shows needles pointing normal to the field direction that rotates counter clockwise starting at direction 12~o'clock. Afterwards, the PMMA was hardened by heating at 140~$^\circ$C under a static magnetic field (see Fig.~\ref{pmma} in Experimental section).

\begin{figure*}
    \centering
    \includegraphics[width=\textwidth]{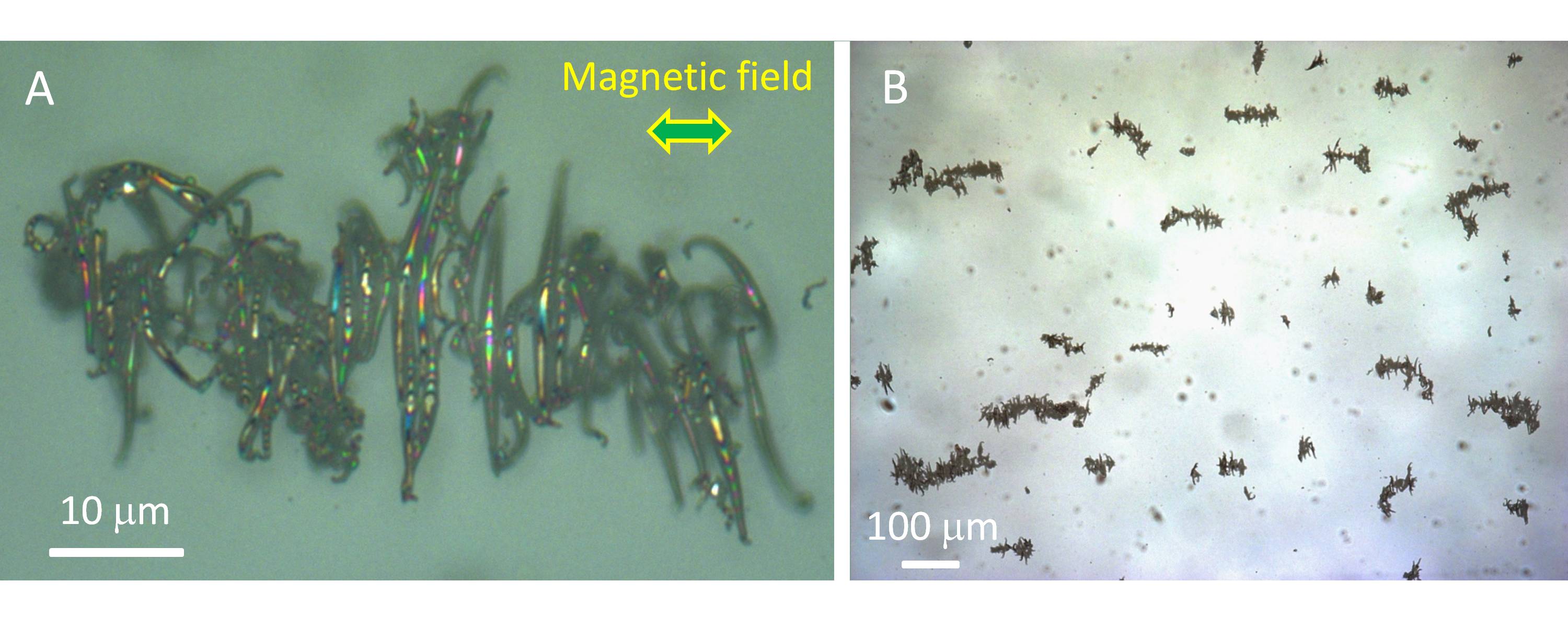}
    \caption{
    (\textbf{A}) High-magnification, and (\textbf{B}) low-magnification optical micrographs of nano-compasses magnetically aligned and bundled in PMMA
    \label{opt}}
\end{figure*}

Figures~\ref{opt}A and \ref{opt}B are typical optical micrographs showing nano-objects bundled together and aligned normal to the field direction encapsulated within the transparent PMMA. This result demonstrates that they function like compasses that point north. However, unlike conventional compasses, the needles point in a direction {\it normal} to the field orientation.

\begin{figure*}
    \centering
    \includegraphics[width=\textwidth]{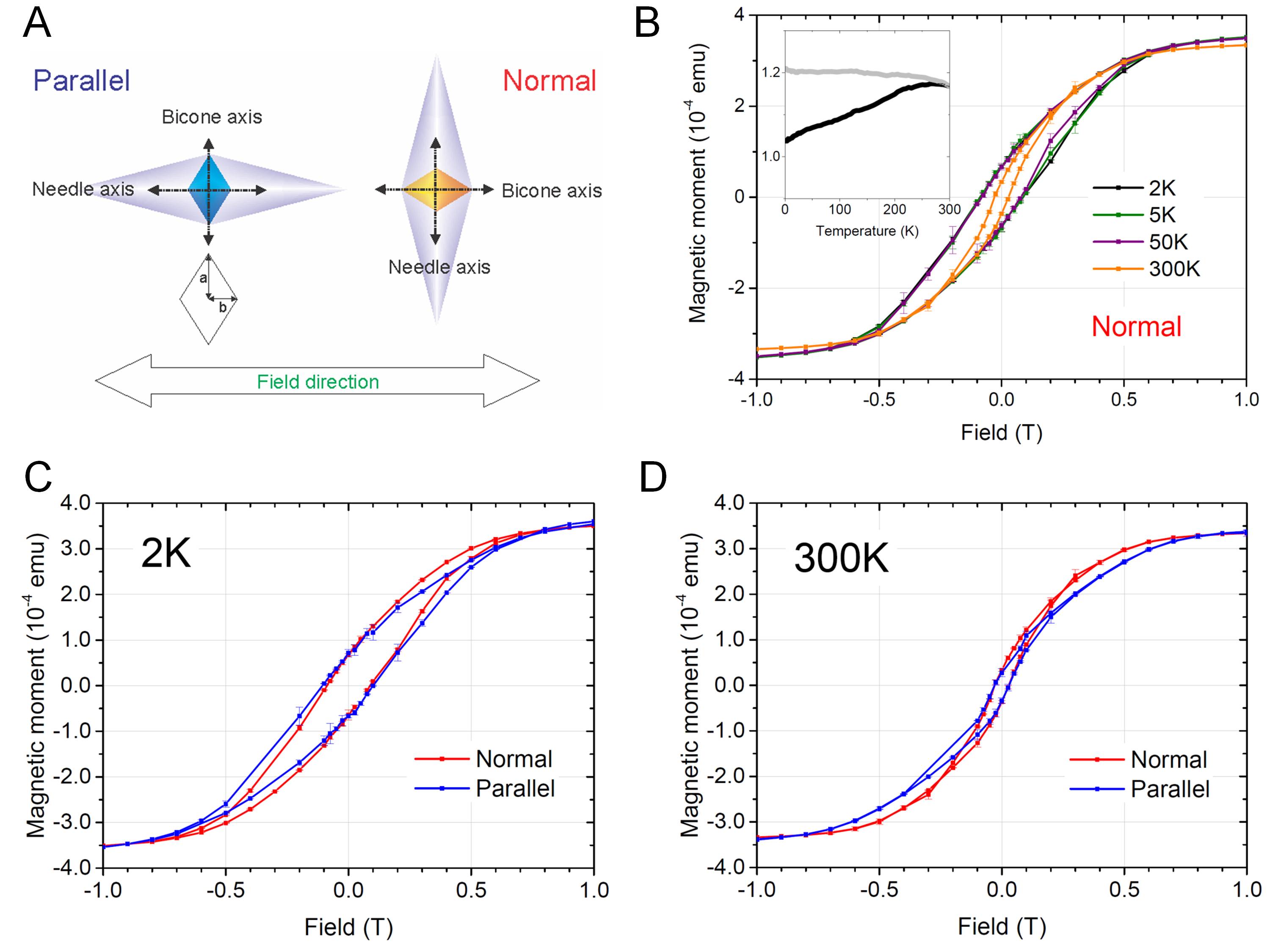}
    \caption{
    (\textbf{A}) Schematics for \textit{Normal} and \textit{Parallel} orientation (\textbf{B}) Magnetization curves recorded in magnetic fields applied normal to the molecular axis at temperatures of 2, 5, 50 and 300~K. Inset: temperature dependence in 100~mT upon field cooling (FC, grey) and after zero field cooling (ZFC, black).
    Magnetization curves collected in magnetic fields applied normal and parallel to the molecular axis at (\textbf{C}) 2~K and (\textbf{D}) 300~K.
    \label{squid}}
\end{figure*}

Further insight into the magnetism can be obtained from bulk magnetization measurements. Figure~\ref{squid}B shows magnetization curves for the aligned nano-compasses collected at temperatures of 2, 5, 50 and 300~K and in magnetic fields up to 1~T applied perpendicular to the molecular axis (see the schematic in Figure~\ref{squid}A). These results show that the nano-compasses are ferromagnetic, with coercive fields as high as $\sim$~78.5--87.5~mT at 2--50~K, reducing to $\sim$27.5~mT at room temperature,
The grain sizes are controlled in a range of 100--1000~nm where the coercive field is optimally enhanced \cite{LUBORSKY61JoAP,HERZER92JoMaMM}, hence the nanocrystalline magnets are magnetically harder than their bulk counterpart.

The inset in Figure~\ref{squid}B shows the magnetic moments versus temperature collected in 100~mT after field cooling (FC) and zero field cooling (ZFC). Upon FC, the net magnetization is observed to increase monotonically as the temperature drops, while the magnetization after ZFC is reduced at low temperatures due to the formation of spin-glass states. This hysteretic behaviour is in agreement with the reduction in coercive field at room temperature. Both are characteristic of superparamagnetism.

Next, we compare magnetization curves for the same sample recorded in magnetic fields applied normal and parallel to the molecular axis, as shown in Figure~\ref{squid}C and D. The data show no differences in saturation magnetization, but marked differences in the magnetic permeability. The steeper slope in the normal direction (red) means that the magnetic permeability in low fields is larger in this direction. This anisotropic permeability leads to anisotropic response of these nano-objects to the applied magnetic field, which could result in the alignment of the molecular axis perpendicular to the field direction, as observed in the optical microscope. The anisotropy could be caused by the core crystal orientation along the molecular axis and more likely by the bicone shape of the core.

As examined previously under a scanning electron microscope \cite{Shiozawa13SR}, the core shape changes from a sphere to a bicone as the particle size becomes greater. For nano-compasses, the cores are bicones that have a base-to-height ratio of $a/b = 1.3$ (see the schematic in Figure~\ref{squid}A).
It is known that when subjected to magnetic fields, finite magnetic objects experience the field produced by magnetic poles on their surface, called the demagnetizing field. The magnetization in low fields is then given as $M = \chi (H_a + H_s ) = \chi (H_a - D M)$ where $\chi$ is the magnetic susceptibility, $H_a$ is the applied magnetic field and $D$ is the demagnetization factor. The experimentally obtained susceptibility is given as $\chi_{exp} = \frac{dM}{dH_a} = \frac{\chi}{1+ \chi D}$, which is reduced with respect to the intrinsic $\chi$.
Now, we approximate a bicone to an ellipsoid. For an ellipsoid with an aspect ratio of $a/b = 1.3$, $D_{\parallel} = 0.2664$ when the magnetic field is applied parallel to the major axis of the ellipsoid and $D_{\perp} = 0.3668$ in transverse orientation \cite{OSBORN45PR}.
This leads to an anisotropy of $\frac{\chi_{\parallel}}{\chi_{\perp}} = \frac{1+ \chi D_{\perp}}{1+ \chi D_{\parallel}} \approx D_{\perp}/D_{\parallel} = 1.377$ for a large $\chi$. Taken into account that the major axis of the core is orthogonal to the needle axis, this explains well the observed anisotropy in bulk magnetization at low fields, Figure~\ref{squid}C and D, while its degree is reduced due to multiple magnetic domains present in the core as well as imperfect alignments of the needles.

Furthermore, the interacting nano-objects are attracted towards one another via confined magnetic fields to make bundles in which the core magnetic dipoles are aligned in a chain structure. When exposed to a magnetic field, each object in a bundle rotates about the needle's axis to maximize the net magnetization along the chain. The total net magnetization at low fields is then along the chain to minimize the demagnetizing field. Thus, these chains could collectively behave as conventional compasses to be aligned along the field, analogous to magnetotactic bacteria in which chains of nano-sized magnetic particles orient and navigate along geomagnetic fields \cite{BLAKEMORE75S}.

\section*{Conclusions}
Here, we have shown that nano-carbon needles encapsulating ferromagnetic cores can be aligned magnetically. Magnetic fields confined near the core, and orthogonally coupled magnetic and structural axes, are responsible for the compass-like behaviour. The controllable and scalable synthesis using the low-cost precursor demonstrated in this study would make the nano-compasses available on a commercial scale and it could serve as a viable alternative to some of the common magnetic materials such as carbonyl iron and ferrous particles that are used in electronic and biomedical applications. Moreover, the graphitic shells are excellent conductors, naturally sustainable against environmental factors, and they could easily be functionalized to become environmentally compatible.

\section*{Experimental Section}
\subsection*{Synthesis}
Ferrocene powder (23.4~mg) was vacuum-sealed in a borosilicate glass ampoule (Outer diameter = 6~mm, Inner diameter = 3~mm, Inner length = 25~mm) and melt at $\sim$ 150~$^\circ$C. The ampoule was placed in a furnace at 650~$^\circ$C, kept for 5~minutes and then slowly cooled to room temperature by switching off the furnace. The powdery soot in the ampoule was collected and prepared for measurements. Most objects observed under an optical microscope were nano-compasses.

\subsection*{Encapsulation in PMMA}
The nano-objects were aligned and encapsulated in PMMA by the following procedure, illustrated in Fig.~\ref{pmma}: as-grown powdery material was dispersed in an ethyl lactate solution of polymethyl methacrylate (PMMA), drop-casted on a glass slide (for optical observations) or a plastic tube (for SQUID measurements), and then placed on a flat neodymium magnet surface along which the magnetic fields are oriented. The sample was placed together with the magnet in an oven at 140~$^\circ$C to cure the PMMA.

\subsection*{Electron microscopy and holography}
For the TEM observations, the powdery soot containing nano-compasses was placed on uniform carbon films on TEM grids.
TEM images and electron holograms were recorded with a Philips-FEI CM 200 electron microscope equipped with a field-emission gun, electrostatic biprism and a charge-coupled-device (CCD) camera. A positive biprism voltage of $\sim$ 100~V, corresponding to an interference-fringe spacing of $\sim$ 2~nm, was used to record holograms with the sample in field-free conditions and the normal objective lens turned off. Further details of holography experimental setup and procedures for phase mapping can be found elsewhere \cite{McCartney07ARoMR}.

\section*{Acknowledgments}
We thank S. Loyer and A. Stangl for technical assistance. This work was supported by the Austrian Science Fund (FWF), projects P621333-N20 and P27769-N20. We also acknowledge use of facilities in the John M. Cowley Center for High Resolution Electron Microscopy at Arizona State University and partial support from DOE Grant DE-FG02-04ER46168.


\end{document}